\documentclass{aastex6}
\turnoffedit

\begin{document}


\title{The two central stars of NGC 1514: \edit1{can they actually be 
related?}\footnote{Based 
on observations obtained at the Canada-France-Hawaii
Telescope (CFHT) which is operated by the National Research Council of Canada,
the Institut National des Sciences de l'Univers of the Centre National de la
Recherche Scientifique of France, and the University of Hawaii.}}


\author{Roberto H. M\'endez\altaffilmark{1}, Rolf-Peter Kudritzki
}
\affil{Institute for Astronomy, 2680 Woodlawn Drive, Honolulu, HI 96822, USA}


\and

\author{Miguel A. Urbaneja}
\affil{Institut f\"ur Astro- und Teilchenphysik, Universit\"at Innsbruck,
Technikerstr. 25/8, A-6020 Innsbruck, Austria}

\altaffiltext{1}{mendez@ifa.hawaii.edu}


\begin{abstract}

The central star of the planetary nebula NGC 1514 is among the visually
brightest central stars in the sky (V=9.5). It has long been known to 
show a composite spectrum, consisting of an A-type star and a much hotter 
star responsible for the ionization of the surrounding nebula. These two 
stars have always been assumed to form a binary system. High-resolution
spectrograms obtained with Espadons at the CFHT on Mauna Kea have allowed
to measure good radial velocities for both stars. They 
differ by 13 $\pm$ 2 km s$^{-1}$. 
The stellar velocities have not changed after 500 days.
We have also estimated the metallicity of the cooler star.
Combining these data with other information available in the literature,
\edit1{we conclude that, unless all the published nebular radial 
velocities are systematically wrong,} the cooler star is just
a chance alignment, and the two stars are not orbiting each other. 
The cooler star cannot have played any role in the formation of NGC 1514.

\end{abstract}


\section{Introduction} \label{sec:intro}

NGC 1514 belongs to a small group of planetary nebulae with A-type
central stars. Obviously in each of these cases a hotter star needs to be 
present, to explain the ionized state of the surrounding nebula. Thus the 
central star of NGC 1514 has always been assumed to be a binary system. 
We will not give an extended historical introduction on NGC 1514; the reader
is directed to a recent paper by Aller et al. (2015). 

That the central star spectrum is composite was shown by Kohoutek (1967),
and confirmed spectroscopically by Greenstein (1972). Greenstein's spectra
did not show any convincing velocity variation, but he did note that 
He {\sc ii} $\lambda$4686 from the hot star showed a mean velocity 
difference of $-7 \pm 5$ km s$^{-1}$ relative to the H and metal lines
of the A-type star. Other observers have given conflicting reports
on radial velocity variations. According to Seaton (1980), this was not
satisfactorily resolved. Since then, it appears that nobody ever tried 
again. We have obtained new high-resolution spectrograms of this 
central star, and report the resulting radial velocities in what 
follows. Section 2 describes the observations. Section 3 explains how the 
velocities were measured, and presents the results. \edit1{In Section 4 we give 
a preliminary estimate of the metallicity of the cooler star. In Section 5 
we discuss all the information we have collected, and Section 6 recapitulates 
the conclusions}.

\section{The Espadons spectrograms}

Espadons is a bench-mounted, high-resolution echelle spectrograph, fiber-fed 
from a Cassegrain unit at the Canada-France-Hawaii Telescope (CFHT) on 
Mauna Kea, Hawaii. Spectrograms of the 9th-magnitude central star of 
NGC 1514 were obtained in Star+Sky mode (R=68000) and normal CCD readout 
mode, with individual exposure times of 1740 seconds. Table 1 gives a list 
of the spectrograms taken for this project. The heliocentric Julian Dates 
correspond to mid-exposure. All the spectrograms were obtained by the CFHT 
staff in Queued Service Observing Mode.

\floattable
\begin{deluxetable*}{cccc}
\tablecaption{Espadons spectrograms used in this work \label{tab:table1}}
\tablehead{
\colhead{Exposure} & \colhead{Date} & \colhead{Heliocentric} & \colhead{Heliocentric} \\
\colhead{Ident Number} & \colhead{} & \colhead{JD (UTC)} & \colhead{RV correction} \\
\colhead{} & \colhead{} & \colhead{(2450000+)} & \colhead{(km s$^{-1}$)}
}
\startdata
1753905 & Nov 05, 2014 & 6967.15142 & +10.847 \\
1753906 &              & 6967.16163 & +10.831 \\
1755174 & Nov 08, 2014 & 6970.07856 & +09.544 \\
1755175 &              & 6970.08877 & +09.522 \\
1756502 & Nov 12, 2014 & 6973.96764 & +07.859 \\
1756503 &              & 6973.97784 & +07.830 \\
1769682 & Dec 20, 2014 & 7011.69743 & -11.020 \\
1769683 &              & 7011.70763 & -11.038 \\
1771245 & Dec 29, 2014 & 7020.69252 & -15.241 \\
1771246 &              & 7020.70272 & -15.262 \\
1895441 & Feb 20, 2016 & 7438.71453 & -29.608 \\
1895442 &              & 7438.73515 & -29.656 \\
\enddata
\end{deluxetable*}

\subsection{Reductions and description}

The spectrograms were fully processed by CFHT using the UPENA software. 
This software calls the Libre-Esprit pipeline (Donati et al. 1997).
We have used the normalized ASCII tables provided by CFHT (filenames 
ending ...in.s), each of which provides the extracted, 
sky-subtracted, wavelength-calibrated spectrum, with rectified continuum,
and with heliocentric radial velocity correction applied. Therefore, 
measurements of wavelengths of any features on these spectra can 
be directly transformed into heliocentric radial velocities.

\subsection{Spectral description of the hotter star}

The spectrum of the cooler star has been sufficiently described in the 
literature (see e.g. Greenstein 1972 and Aller et al. 2015). The high 
spectral resolution and high signal-to-noise ratio of the Espadons 
spectra permit a good description of the hot star features. The 
following are clearly visible in absorption (all wavelengths in \AA): 
He {\sc ii} 4200, 4541, 4686, 5411; N {\sc v} 4603, 4619; O{\sc v} 5114. 
There are also narrow emissions of C {\sc iv} 5801, 5811, N {\sc iv} 4057, 
N {\sc v} 4944, O {\sc v} 4930, and O {\sc vi} 5290. These 
high-ionization C, N, O
features, together with the absence of C {\sc iii} 5696, N {\sc iii} 
4634, 4640, 4641, and O {\sc iii} 5592, indicate a very high surface
temperature of this star, as expected from the 90,000 K estimated
by Aller et al. (2015). The strength of the Balmer absorption lines from 
the A-type star makes it very hard to decide if the hotter star is 
H-deficient. There is no evidence of strong mass loss in the optical 
spectrum. If Aller et al. (2015) are correct, and some H is present
on the surface, then the spectral type, in the system of M\'endez (1991), 
is O(H).

Figures 1 and 2 show the result of co-adding 
the 12 Espadons spectrograms. For brevity we show only 
the most interesting spectral features of the hot star.
The sharpness of the C, N, O features indicates that $v$ sin $i$ 
\edit1{and the macroturbulent velocity} must be rather 
low, and that no large-amplitude radial velocity variations can be expected
in the available sample.
A more detailed spectral analysis is deferred to future work.

\section{Radial velocity measurements}

Because of the composite nature of the spectrum, we decided to measure
spectral features individually, rejecting any features where 
contamination from the other star can be expected. For example,
some He {\sc ii} absorption profiles are distorted by other 
absorptions from the A-type star, or other features, and were not used.
N {\sc v} 4619 was rejected for the same reason.
Table 2 lists the adopted wavelengths of all the spectral features 
measured for radial velocity on the individual spectra. We selected 
only the highest quality, symmetric profiles.

\floattable
\begin{deluxetable*}{cclc}
\tablecaption{Spectral features measured for radial velocity \label{tab:table2}}
\tablehead{
\colhead{Star} & \colhead{Element} & \colhead{Wavelength (\AA)} & \colhead{} \\
}
\startdata
Cooler  & Mg {\sc ii} & 4481.228 & ab \\
Cooler  & Fe {\sc ii} & 4508.27  & ab \\
Cooler  & Fe {\sc ii} & 5018.434 & ab \\
Cooler  & Fe {\sc ii} & 5316.69  & ab \\
Cooler  & Si {\sc ii} & 6347.091 & ab \\
Hotter  & N  {\sc iv} & 4057.759 & em \\
Hotter  & N  {\sc v}  & 4603.73  & ab \\
Hotter  & O  {\sc v}  & 5114.07  & ab \\
Hotter  & C  {\sc iv} & 5801.33  & em \\
Hotter  & C  {\sc iv} & 5811.98  & em \\
\enddata
\end{deluxetable*}

Radial velocities were measured in the following way. 
First, it was necessary to create a new ASCII table for each spectrum,
with equally spaced wavelengths always taken from a unique prespecified set. 
The intensities 
corresponding to each wavelength in the prespecified set were obtained 
from the original spectrum by a standard interpolation routine.

The next step was to co-add all the spectra; this co-added spectrum is the 
source of the plots shown in Figures 1 and 2. Then, on the co-added spectrum,
for every spectral feature in Table 2 a Gaussian was fitted, and
the peak of the Gaussian was used to calculate the heliocentric
radial velocity, in km s{$^{-1}$}, to be used as reference. The goodness 
of the Gaussian fit was always checked interactively.

Finally, the individual velocities for each feature and for each individual
spectrum were calculated by cross-correlating against the corresponding 
feature in the co-added spectrum.
Tables 3 and 4 give the results for the cooler and hotter star, respectively.

\floattable
\begin{deluxetable*}{ccccccc}
\tablecaption{Heliocentric radial velocities of the cooler star in km s{$^{-1}$}\label{tab:table3}}
\tablehead{
\colhead{Exposure} & \colhead{Mg II} & \colhead{Fe II} & \colhead{Fe II} & \colhead{Fe II} & \colhead{Si II} & \colhead{Heliocentric} \\
\colhead{Ident} & \colhead{4481} & \colhead{4508} & \colhead{5018} & \colhead{5316} & \colhead{6347} & \colhead{JD (UTC)} \\
\colhead{Number} & \colhead{} & \colhead{} & \colhead{} & \colhead{} & \colhead{} & \colhead{(2450000+)}
}
\startdata 
1753905 &  45.0 & 46.2 & 48.8 & 43.9 & 41.6 &  6967.15142 \\
1753906 &  44.4 & 44.7 & 48.3 & 43.6 & 42.9 &  6967.16163 \\
1755174 &  43.8 & 43.8 & 45.6 & 41.4 & 40.5 &  6970.07856 \\
1755175 &  45.2 & 45.1 & 46.3 & 39.8 & 41.5 &  6970.08877 \\
1756502 &  44.9 & 47.2 & 47.3 & 41.8 & 43.3 &  6973.96764 \\
1756503 &  45.1 & 45.5 & 48.1 & 41.4 & 43.0 &  6973.97784 \\
1769682 &  44.2 & 42.7 & 47.6 & 41.0 & 42.8 &  7011.69743 \\
1769683 &  43.6 & 41.6 & 46.3 & 39.9 & 43.1 &  7011.70763 \\
1771245 &  44.4 & 43.9 & 46.5 & 40.8 & 44.5 &  7020.69252 \\
1771246 &  44.0 & 45.7 & 46.1 & 41.0 & 42.9 &  7020.70272 \\
1895441 &  44.6 & 44.9 & 47.4 & 41.4 & 42.9 &  7438.71453 \\
1895442 &  45.0 & 44.7 & 47.6 & 41.8 & 44.6 &  7438.73515 \\
\enddata
\end{deluxetable*}

\floattable
\begin{deluxetable*}{ccccccc}
\tablecaption{Heliocentric radial velocities of the hotter star in km s{$^{-1}$}\label{tab:table4}}
\tablehead{
\colhead{Exposure} & \colhead{N IV} & \colhead{N V} & \colhead{O V} & \colhead{C IV} & \colhead{C IV} & \colhead{Heliocentric} \\
\colhead{Ident} & \colhead{4057} & \colhead{4603} & \colhead{5114} & \colhead{5801} & \colhead{5811} & \colhead{JD (UTC)} \\
\colhead{Number} & \colhead{} & \colhead{} & \colhead{} & \colhead{} & \colhead{} & \colhead{(2450000+)}
}
\startdata  
1753905 &   59.0 & 57.3 & 57.8 & 55.4 & 57.6 &   6967.15142 \\
1753906 &   59.0 & 57.5 & 57.7 & 54.9 & 57.1 &   6967.16163 \\
1755174 &   54.8 & 58.0 & 57.9 & 55.1 & 56.7 &   6970.07856 \\
1755175 &   58.2 & 57.9 & 58.8 & 55.1 & 56.7 &   6970.08877 \\
1756502 &   58.1 & 56.7 & 58.3 & 55.0 & 56.7 &   6973.96764 \\
1756503 &   60.3 & 56.8 & 57.5 & 55.5 & 56.5 &   6973.97784 \\
1769682 &   56.7 & 56.8 & 57.8 & 55.9 & 57.0 &   7011.69743 \\
1769683 &   59.6 & 56.9 & 58.1 & 55.5 & 57.6 &   7011.70763 \\
1771245 &   56.1 & 56.9 & 56.8 & 55.4 & 57.3 &   7020.69252 \\
1771246 &   58.6 & 56.9 & 57.2 & 55.1 & 56.9 &   7020.70272 \\
1895441 &   57.6 & 56.6 & 56.5 & 54.5 & 56.7 &   7438.71453 \\
1895442 &   57.2 & 56.2 & 56.5 & 54.5 & 56.0 &   7438.73515 \\
\enddata
\end{deluxetable*}

\section{The metallicity of the cooler star}

Armed with our high-resolution co-added spectrogram, we can study
the metallicity of the cooler star. As 
discussed by Aller et al. (2015), for the observed combination of 
$T_{\rm eff}$ = 9850 K and log $g$ = 3.5, we have two alternatives: 
a horizontal branch star, or a more 
massive star evolving away from the main sequence. But the horizontal
branch star, to be hot enough, must have a low metallicity. For example,
the evolutionary tracks used by Aller et al.\ were calculated with 
$Z=0.006$. Another way of seeing this would be to look at Figure 2.2 in
Ashman \& Zepf (1998), where they compare color-magnitude diagrams of
two globular clusters with different metallicities ([Fe/H] = -1.7 and
-0.6). The more metal-poor cluster shows a well-populated horizontal 
branch at (B-V)=0, while the other has nothing at that position.

Therefore, a sufficiently accurate measurement of metallicity can be 
used to test if the cooler star can be a horizontal branch star. 
To obtain the spectrum of the cooler star, we 
adopted the ratio of fluxes shown in Figure 5 of Aller et al. (2015),
subtracted the contaminating light from the hotter star, and renormalized.
The spectrum obtained in this way was compared with synthetic normalized 
spectra calculated with line-blanketed model atmospheres and very detailed
NLTE line formation calculations (Przybilla et al. 2006). Similar models
have been successfully used by Kudritzki et al. (2008, 2012) to study
medium-resolution spectra of supergiant stars in nearby galaxies.

Synthetic spectra were calculated for log $g$ = 3.5; a pair of 
temperatures, $T_{\rm eff}$ = 9700 K and 10,000 K, bracketing the temperature 
determination of Aller et al. (2015); and a pair of metallicities:
[$Z$]=log($Z/Z\odot$) = 0.0 and -0.6. For a definition of this [$Z$]
we refer e.g. to Kudritzki et al. (2008).

To apply these synthetic spectra to a comparison with our Espadons 
spectrogram, the synthetic spectra were degraded to the
instrumental resolution, 
and further broadened with a v sin i of 50 km s{$^{-1}$, obtained from the
renormalized profiles of a few blend-free absorption lines in the spectrum
of the cooler star. The results of the comparison of observed versus 
synthetic spectra can be appreciated in Figures 3 and 4. The observed
profiles of metal lines fall, on average, in between the two corresponding
synthetic profiles. We conclude that for the cooler star [$Z$] is 
approximately -0.3. We defer 
a more careful determination to future work, where we expect also to measure
abundances in the hotter star. For the moment, Figs. 3 and 4 confirm that
the metallicity of the cooler star is too high for a horizontal branch star.
Therefore, it must be a more massive \edit1{(around 3 M$\odot$)}, and more 
luminous star, at a distance 
of \edit1{at least 400} pc (Aller et al. 2015). We hope that very soon 
the Gaia mission will be able to test this prediction.


\section{Discussion}

The average velocity of the cooler star in Table 3 is 
44 $\pm$ 2 km s{$^{-1}$}. The average velocity of the hotter star 
in Table 4 is 57 $\pm$ 1 km s{$^{-1}$}. The small differences between 
the average velocities from the different lines are probably 
due to small errors in the laboratory wavelengths.
The velocities of 
the two stars differ significantly, by 13 $\pm$ 2 km s{$^{-1}$}. 
None of the two stars have shown substantial velocity variations.

Assume the two stars are orbiting each other. The different velocities
indicate that we cannot be observing the orbits pole-on. Then the lack of
variations of the order of 10 km s{$^{-1}$} in almost two years would 
require a long orbital period. The results in Tables 3 and 4 can be used to 
reject all previous claims or suggestions of high-amplitude, short-period 
radial velocity variations 
(e.g. as discussed by Muthu \& Anandarao 2003 in their section 5). 
\edit1{This implies that the cooler star cannot have played any direct role
in the formation of NGC 1514, e.g. through a common envelope episode.}

There is some additional information to be extracted from the existing 
literature.


\subsection{Radial velocity of the cooler star}

The only previous study at high spectral resolution is by Greenstein (1972).
There are coud\'e spectrograms from 1949 to 1971. The average radial 
velocity (47.6 $\pm$ 1.6 km s{$^{-1}$})  
looks quite in agreement with the values reported in Table 3. The 
Greenstein velocities are not directly comparable to those in Table 3, 
because he included several lines we have rejected, like the Balmer lines;
did not give the velocities for each spectral feature
separately; nor did he report what 
laboratory wavelengths he used. But given the dispersion of values 
listed in his Table 1, and his conclusion (no sign of velocity 
variation, in his own words), it is reasonable to
conclude that there is no reliable evidence of variations in the velocity
of the cooler star in six different epochs from 1949 to 2016.


\subsection{Radial velocity of the hotter star}

In Greenstein's paper the velocity of He {\sc ii} 4686 is mentioned to
be more negative than that of the Balmer lines and the metals by 7 $\pm$ 5
km s{$^{-1}$}. He did not elaborate on this, so probably he did not 
consider the difference to be significant. 

The velocity we measured on the co-added Espadons spectrum for He {\sc ii} 
4686 is 52 km s{$^{-1}$}. We did not include velocities from this line in 
Table 4 because the profile is affected by what appears to be an 
unidentified emission line at 4688 \AA \ (see Figure 1). Another reason 
to avoid this line as a radial velocity indicator is that it is among the 
first, in the visible spectral range, to suffer the effects of a stellar 
wind. Even a very incipient P Cygni-like profile would induce a shortward 
wavelength displacement in the observed absorption (Kudritzki et al.\ 2006).

\edit1{Continuous monitoring for at least a decade would be required to
verify if Greenstein's data were implying long-period variability
of the hotter star's velocity.}

\subsection{Radial velocity of the nebula}

According to Schneider et al. (1983), the heliocentric radial velocity 
of NGC 1514 is 60 $\pm$ 4 km s{$^{-1}$} (an average of six different 
determinations, each with a reported uncertainty of between 3 and 9 
km s{$^{-1}$}). This is perfectly compatible with the velocity of 
the hotter star in Table 4.

Two velocities in Schneider et al. are discrepant: one substantially
above the average, and one below it. The one above can be ignored, 
because of its lower accuracy. The one below is more interesting: it is
41 $\pm$ 5 km s{$^{-1}$, from Greenstein (1972). There is a way of 
understanding 
the discrepancy. Greenstein reports that the spectrograms he used to 
measure the nebular velocity were taken with the slit located one arc 
minute south of the central star. There is a spatiokinematic study of 
NGC 1514 by Muthu \& Anandarao (2003), made with a Fabry-Perot spectrometer.
Unfortunately they did not attempt to measure the absolute radial velocity;
but they do show velocity channel maps (their Figure 4). In this figure
we find that slightly less than one arc minute south of the central star 
there is a concentration of gas with a velocity of -13 km s{$^{-1}$.
There is no corresponding gas at the opposite velocity (+13).
That is conceivably the reason why Greenstein's nebular velocity
was more negative.

Anyway, it is fair to say that systematic errors in the published
nebular velocities cannot be completely discarded. 
Integral field spectroscopy of NGC 1514 at high spectral 
resolution would permit to accurately map the whole 
velocity field, and produce a more reliable number for the systemic 
radial velocity of this planetary nebula. 

\subsection{Binary system or chance alignment?}

Since the radial velocities of the two central stars differ, 
\edit1{and the nebular velocity agrees with that of the hotter star,}
there is a distinct possibility of chance alignment 
\edit1{with the cooler star}. This
would be extremely surprising, because Ciardullo et al. (1999) were 
unable to resolve the pair on Hubble Space Telescope (HST) images. 
The probability of such a perfect chance alignment is extremely 
low, of the order of $10^{-8}$ (using an area of 0.01 square arc 
second, and an estimated surface density of 10 stars of 10th visual 
magnitude per square degree at the galactic latitude of NGC 1514, 
which is 15 degrees). See e.g. Figure 4 of Bahcall \& Soneira (1980). 

\edit1{
In dealing with such a low probability, every possible alternative, 
however unlikely, must be considered. Our anonymous referee has kindly 
provided a few.} 

\edit1{
Let us first consider if the velocity of the hotter star could be higher
because of gravitational redshift. This would require its log $g$ to be 
around 7.5. But this replaces one problem with another: since the hotter 
star is rather bright, its 
distance from us would have to be much less than 50 pc, ruling out any
association with the cooler star.}

\edit1{
Next, imagine that the velocity of the cooler star is different because
of a recoil effect produced in a binary system when NGC 1514 was formed. 
The problem here is that, in order to produce a 
substantial recoil effect, the orbital period would have to be longer than
the time scale for nebular ejection. But for such long orbital periods, the 
orbital velocities would not be enough to explain the observed velocity 
difference between the two stars (a few examples are given in Table 5, to
be explained below).}

\edit1{
Finally, under what conditions can the two central stars of NGC 1514 
be members of a binary system? In Section 4 we concluded that the 
cooler star must be rather massive, around 3 M$\odot$. Assume a typical
central star mass of 0.6 M$\odot$ for the hotter star. If these two stars 
are orbiting each other, the nebular velocity must be close to the velocity
of the cooler star. In other words, all the nebular measurements discussed 
in the previous subsection would have to be systematically wrong.
}

Consider the value of the sum of the radial velocity semiamplitudes as a 
function of the orbital period. 
Table 5 shows the values of
the sum of the orbital semimajor axes (a$_1$ + a$_2$) and the sum of the 
radial velocity semiamplitudes (K$_1$ + K$_2$) as functions of the 
orbital period, using Newton's version of Kepler's 3rd law, with a 
total mass of \edit1{3.6} M$\odot$, and using the relation between 
semiamplitude, orbital period, semimajor axis and eccentricity.
Taking an orbital inclination of 53 degrees and an eccentricity e=0.6,
the factors sin $i$ and $(1-e^2)^{0.5}$ cancel each other.
\edit1{Unfavorable (smaller) values of inclination} would produce
smaller numbers in the third column of Table 5. \edit1{An orbital 
eccentricity $e$=0.7 would increase those third-column numbers by a 
factor 1.4. But on the other hand, most of the time, the stars would
be far from periastron, and the velocity difference between
the two stars would be actually smaller.}

\floattable
\begin{deluxetable}{rrr}
\tablecaption{Radial velocity amplitudes for different 
periods \label{tab:table5}}
\tablehead{
\colhead{Period} & \colhead{a$_1$ + a$_2$} & \colhead{K$_1$ + K$_2$}  \\
\colhead{(years)} & \colhead{(R$\odot$)} & \colhead{(km s{$^{-1}$)}}
}
\startdata
   1   &   329  &   46   \\
  10   &  1529  &   21   \\
  30   &  3180  &   15   \\
 100   &  7096  &   10   \\
\enddata
\end{deluxetable}

Table 5 \edit1{suggests an upper limit of about 50} years on the 
possible orbital period. Any longer period would require a value of 
K$_1$ + K$_2$ smaller than the observed velocity difference of 13 
km s{$^{-1}$. 
\edit1{It will not take more than a few decades to decide
conclusively if these two stars can actually be related.}

\subsection{Radial velocities: summary and consequences}

The Espadons spectrograms show that the radial velocities of the two central 
stars of NGC 1514 are clearly different. The constancy of these high-quality
velocities over a time of two years permits to reject all previous claims or 
suggestions of substantial short-term changes in the stellar velocity. 
\edit1{Even if they orbit each other, the implied separation 
between the two stars precludes any past evolutionary interaction like a 
common-envelope episode.} For all practical purposes, these two stars are 
\edit1{evolutionarily} unrelated. 
\edit1{On the other hand, if they orbit each other, 
arguing from Kepler's 3rd law, given the observed velocity difference,
we do not expect an orbital period longer than about 50 years.}


\edit1{If new measurements of the nebular radial velocity confirm the
values reported in the literature, which agree with the radial velocity
of the hotter star,} we will have to conclude that, unlikely as it may seem, 
the cooler star is a chance alignment. There is the problem that 
Ciardullo et al. (1999) could not resolve the two 
stars on their HST images. If there is a significant difference in the proper 
motions, then we would expect that sooner or later the two stars will 
become resolvable. In 1791 William Herschel 
reported that the star was perfectly in the center of NGC 1514 (see 
e.g.\ Greenstein 1972), and it still is today. 
\edit1{The proper motion of the 
cooler star, measured by the Hipparcos satellite, is 8.4 milliarcsec/yr.
This amounts to about 1.7 arcsec since the time of Herschel. On the other 
hand, the galactic longitude of NGC 1514 is 165 degrees, i.e., almost in the 
anticenter direction; so it would not be terribly surprising to find the
two stars, even if at somewhat different distances from us, having similar 
proper motions. What is required now is a new observation with the highest 
possible angular resolution, which we expect to make in the near future.}

The results reported here do not preclude the existence of low-amplitude
velocity variations; both stars could have planetary systems or low-mass 
companions. Only future measurements of comparable or higher quality will 
tell.

\section{Recapitulation}

The Espadons spectrograms have permitted a much better spectral description 
of the hotter central star in NGC 1514 than previously available. On these 
new spectra it has
been possible to measure the radial velocities of the two central stars.
They are different, by about 13 $\pm$ 2 km s{$^{-1}$. No evidence of
radial velocity variations has been found in either star. 
\edit1{The metallicity of the cooler star is incompatible with its
belonging to the horizontal branch; the cooler star must be more massive, 
more luminous, and more distant. We hope the Gaia mission will soon test our
prediction of a distance of at least 400 pc for the cooler star.}


\edit1{The published nebular velocities agree with the hotter star's 
velocity, but might be somewhat unreliable. If the two central stars 
of NGC 1514 are a binary system, then we expect the nebular velocity 
to agree more with the cooler star's velocity. If future measurements of 
the nebular velocity confirm its similarity to the hotter star's 
velocity, and if no evidence of mutual orbital motion becomes apparent
in the next few decades, we will have to conclude that the cooler 
star is a chance alignment. In any case,} the cooler star cannot
have played any direct role in the formation of NGC 1514.

{\bf Acknowledgements}. We would like to express our gratitude to the staff 
of the Canada-France-Hawaii Telescope, for performing the necessary 
observations in Queued Service Observing Mode, and for providing
beautifully reduced spectrograms. 
\edit1{We thank the anomymous referee for 
making several valuable suggestions.}

\newpage

\begin{figure*}[ht!]
\figurenum{1}
\plotone{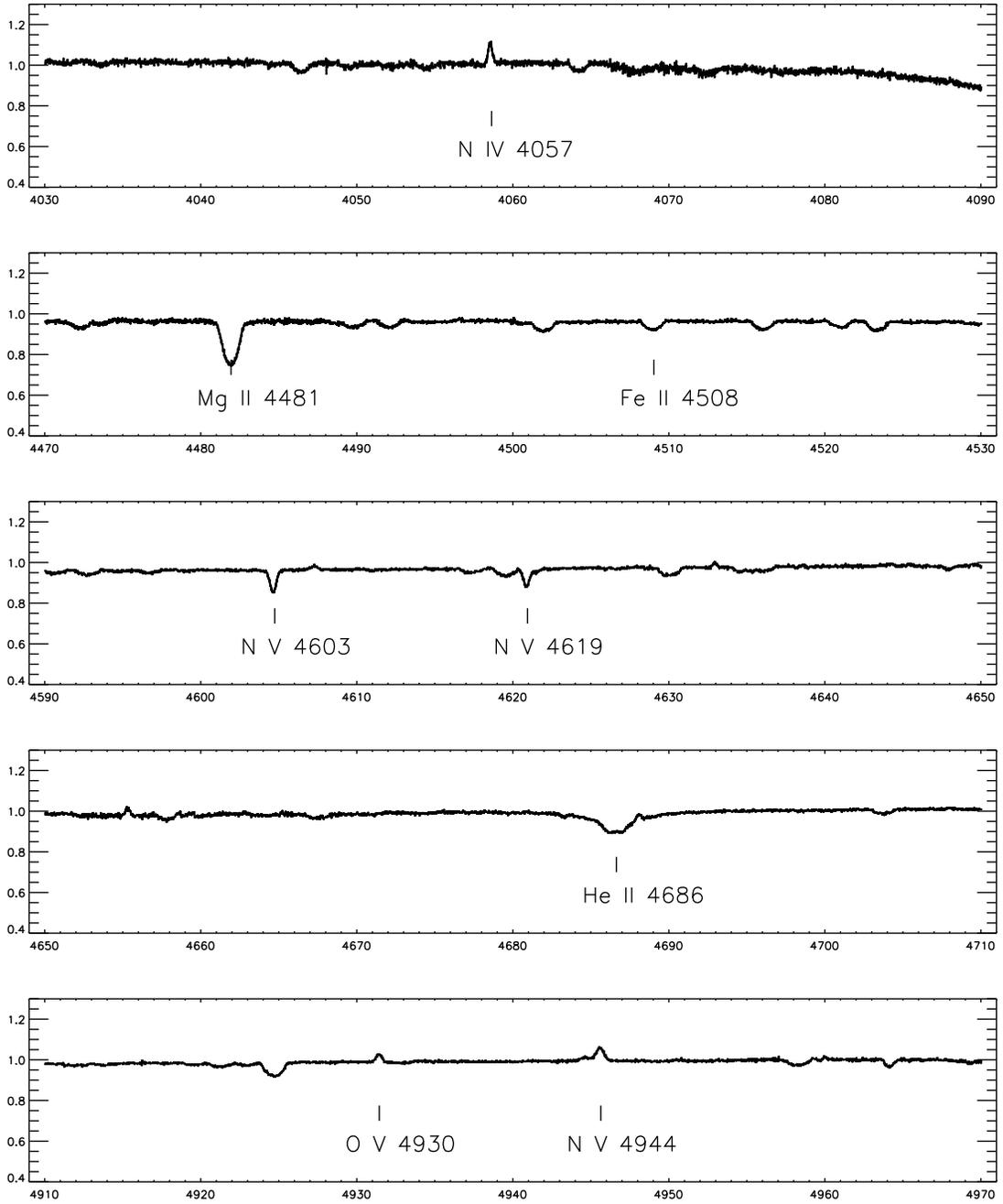}
\caption{Selected features in the composite spectrum of the central 
stars of NGC 1514. The 12 Espadons spectra have been co-added. No attempt 
was made to optimize the continuum rectification. All wavelengths are
in \AA.\label{fig:1}}
\end{figure*}

\newpage

\begin{figure*}[ht!]
\figurenum{2}
\plotone{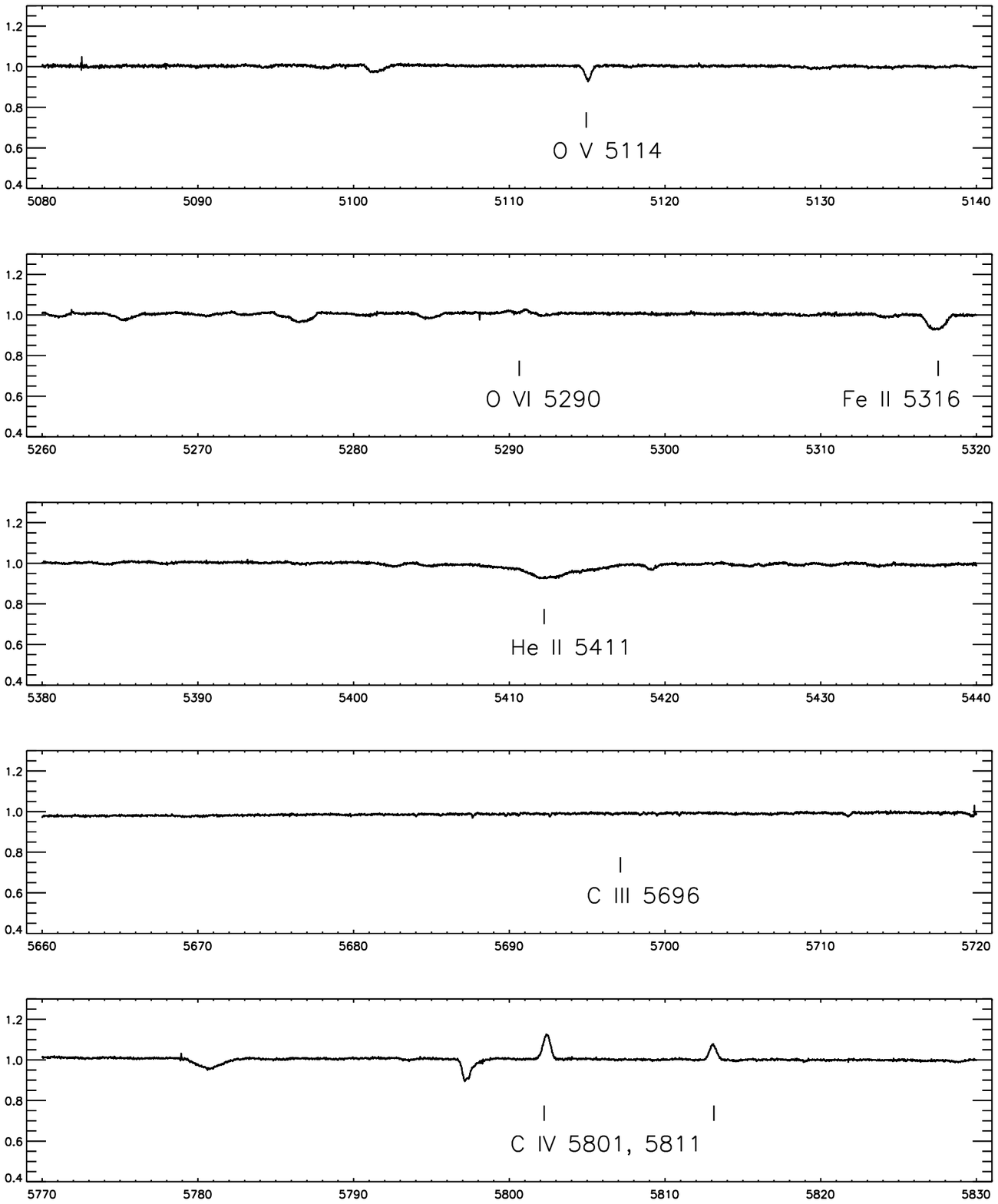}
\caption{Selected features in the composite spectrum of the central 
stars of NGC 1514. The 12 Espadons spectra have been co-added. No attempt 
was made to optimize the continuum rectification. All wavelengths are
in \AA.\label{fig:2}}
\end{figure*}

\newpage

\begin{figure*}[ht!]
\figurenum{3}
\includegraphics[width=14cm, angle=90]{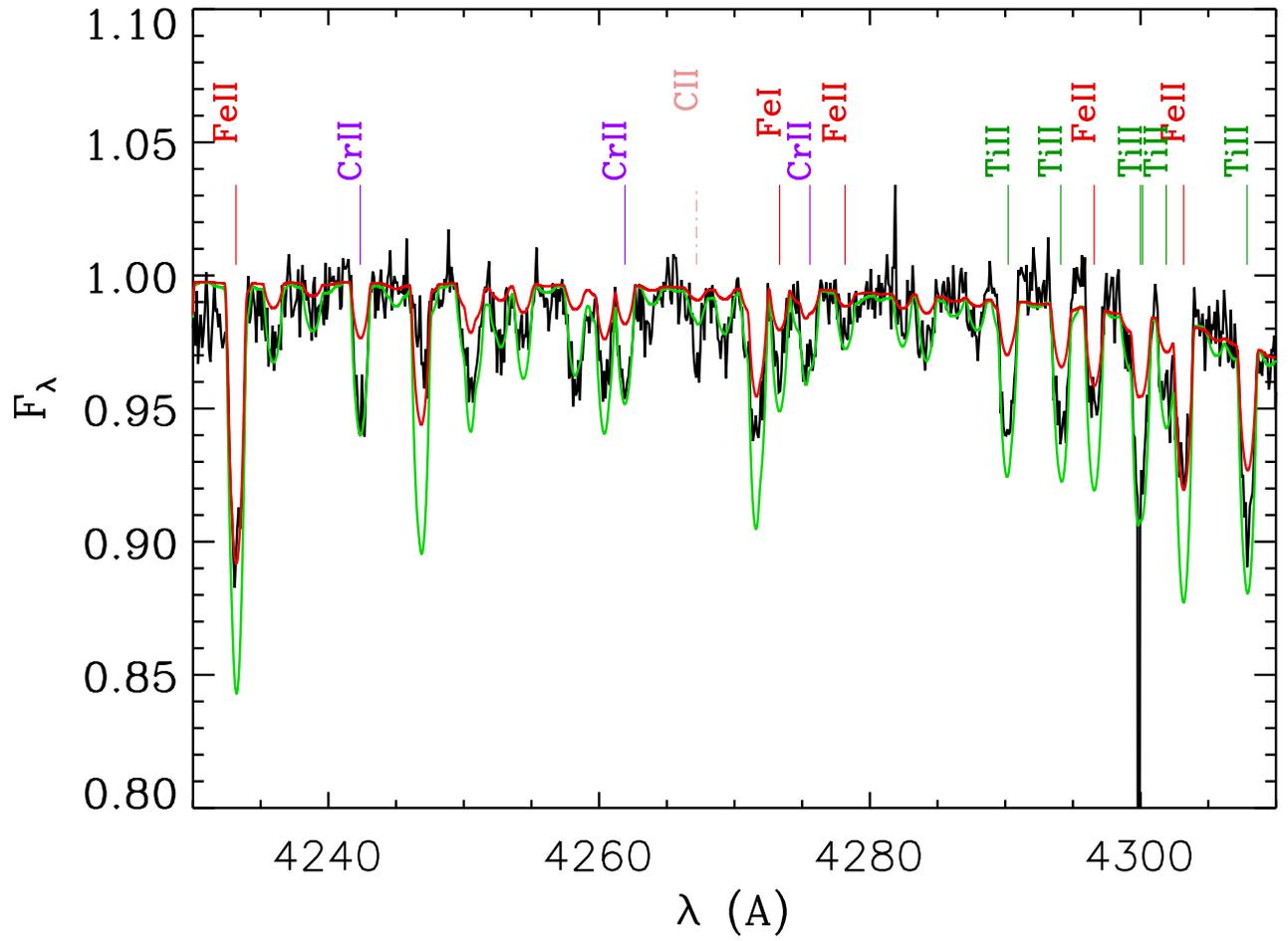}
\caption{Comparison of the observed spectrum of the cooler star versus 
synthetic spectra for log $g$ = 3.5, $T_{\rm eff}$ = 9700 K, and two
metallicities, [$Z$]=0.0 (green) and -0.6 (red). On average, we estimate
the cooler star has a [$Z$] of about -0.3.\label{fig:3}}
\end{figure*}

\newpage

\begin{figure*}[ht!]
\figurenum{4}
\includegraphics[width=14cm, angle=90]{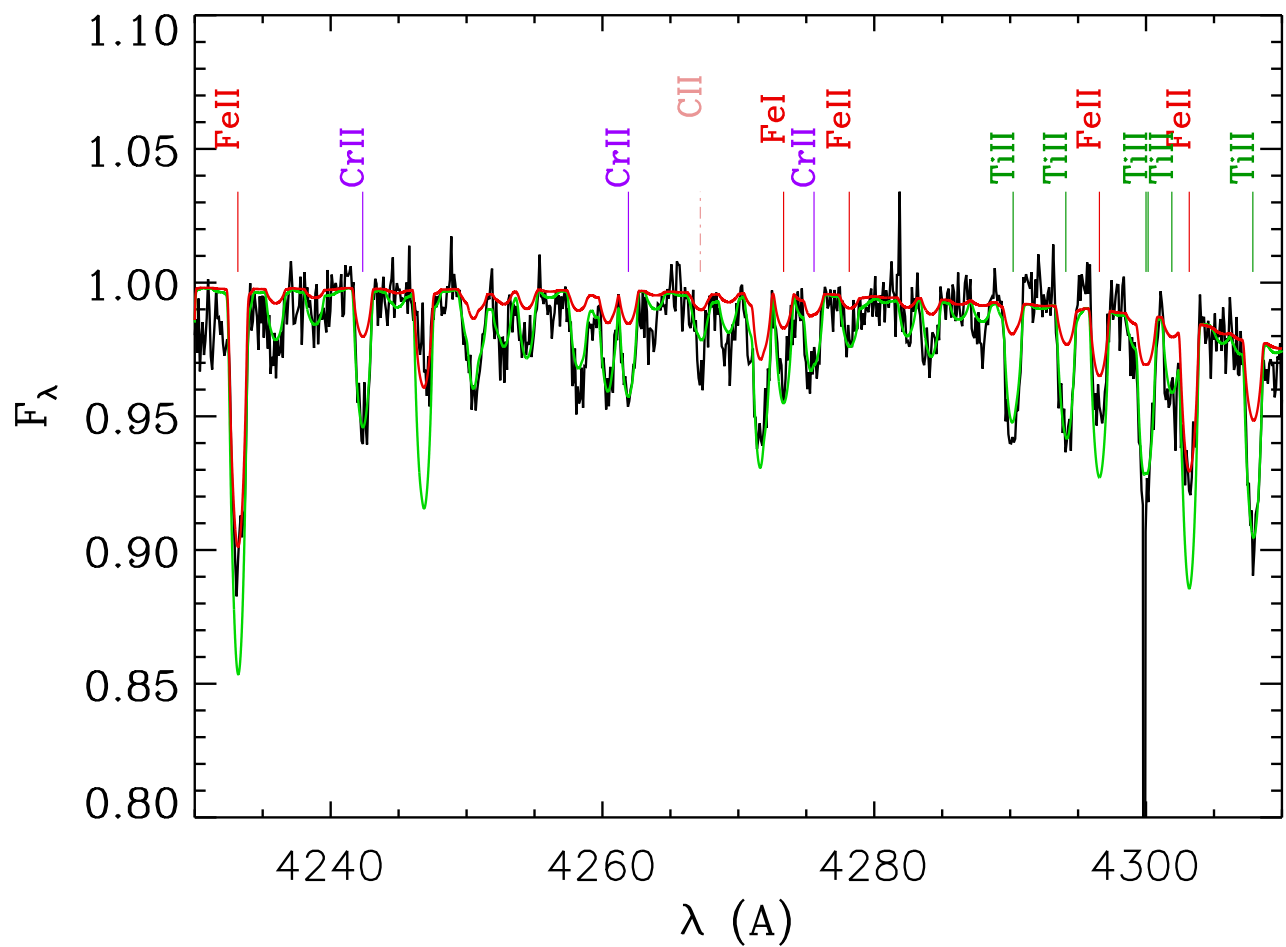}
\caption{Same as Figure 3, but with $T_{\rm eff}$ = 10,000 K. \label{fig:4}}
\end{figure*}



\newpage

\begin{thebibliography}{}

\bibitem[Aller et al.(2015)]{Aller2015} Aller, A., Montesinos, B., 
         Miranda, L.F., et al.\ 2015, MNRAS, 448, 2822

\bibitem[Ashman \& Zepf(1998)]{AsZe1998} Ashman, K.M., \& Zepf, S.E.\
         1998, Globular Cluster Systems, Cambridge Astrophysics Series,
         30, p.\ 9 (Cambridge University Press)

\bibitem[Bahcall \& Soneira(1980)]{BahSon1980} Bahcall, J.N., \& Soneira,
         R.M.\ 1980, ApJS, 44, 73

\bibitem[Ciardullo et al.(1999)]{Ciard1999} Ciardullo, R., Bond, H.E.,
         Sipior, M.S., et al.\ 1999, AJ, 118, 488

\bibitem[Donati et al.(1997)]{Dona1997} Donati, J.-F., Semel, M.,
         Carter, B.D., et al.\ 1997, MNRAS, 291, 658

\bibitem[Greenstein(1972]{Gree1972} Greenstein, J.L.\ 1972, ApJ, 173, 367

\bibitem[Kohoutek(1967)]{Koho1967} Kohoutek, L.\ 1967, BAICz, 18, 103 

\bibitem[Kudritzki et al.(2008)]{Kud2008} Kudritzki, R.P., Urbaneja, M.A., 
         Bresolin, F., et al.\ 2008, ApJ, 681, 269

\bibitem[Kudritzki et al.(2012)]{Kud2012} Kudritzki, R.P., Urbaneja, M.A.,
         Gazak, Z., et al.\ 2012, ApJ, 747:15 

\bibitem[Kudritzki et al.(2006)]{Kud2006} Kudritzki, R.P., Urbaneja, M.A., 
         \& Puls, J.\ 2006, in IAU Symp 234, 
         eds. M.J.Barlow \& R.H.M\'endez, p. 119

\bibitem[Mendez(1991)]{Mend1991} M\'endez, R.H.\ 1991, in IAU Symp 145,
         eds. G.Michaud \& A.Tutukov, p.\ 375

\bibitem[Muthu \& Anandarao(2003)]{Mut2003} Muthu, C., \& 
         Anandarao, B.G.\ 2003, AJ, 126, 2963

\bibitem[Przybilla et al.(2006)]{Przy2006} Przybilla, N., Butler, K., 
         Becker, S.R., et al.\ 2006, A\&A, 445, 1099

\bibitem[Schneider et al.(1983)]{Sch1983} Schneider, S.E., Terzian, Y.,
         Purgathofer, A., \& Perinotto, M.\ 1983, ApJS, 52, 399

\bibitem[Seaton(1980)]{Seat1980} Seaton, M.J.\ 1980, QJRAS, 21, 229

\end{thebibliography}
\end{document}